\documentclass[aps, prl, reprint, showpacs, showkeys, onecolumn, 11pt, superscriptaddress]{revtex4-1}
\usepackage{graphicx} 

\begin{document}

\bibliographystyle{apsrev4-1}

\title{Transform-limited single photons from a single quantum dot}

\author{Andreas V. Kuhlmann}
\affiliation{Department of Physics, University of Basel, Klingelbergstrasse 82, CH-4056 Basel, Switzerland}

\author{Jonathan H. Prechtel}
\affiliation{Department of Physics, University of Basel, Klingelbergstrasse 82, CH-4056 Basel, Switzerland}

\author{Julien Houel}
\affiliation{Department of Physics, University of Basel, Klingelbergstrasse 82, CH-4056 Basel, Switzerland}
\affiliation{Institut Lumi\`{e}re Mati\`{e}re (ILM), UMR5306 Universit\'{e} Lyon 1/CNRS, Universit\'{e} de Lyon, 69622 Villeurbanne Cedex, France}

\author{Arne Ludwig}
\affiliation{Lehrstuhl f\"{u}r Angewandte Festk\"{o}rperphysik, Ruhr-Universit\"{a}t Bochum, D-44780 Bochum, Germany}

\author{Dirk Reuter}
\affiliation{Lehrstuhl f\"{u}r Angewandte Festk\"{o}rperphysik, Ruhr-Universit\"{a}t Bochum, D-44780 Bochum, Germany}
\affiliation{Department Physik, Universit\"{a}t Paderborn, Warburger Strasse 100, D-33098 Paderborn, Germany}

\author{Andreas D. Wieck}
\affiliation{Lehrstuhl f\"{u}r Angewandte Festk\"{o}rperphysik, Ruhr-Universit\"{a}t Bochum, D-44780 Bochum, Germany}

\author{Richard J. Warburton}
\affiliation{Department of Physics, University of Basel, Klingelbergstrasse 82, CH-4056 Basel, Switzerland}

\date{\today}

\begin{abstract}
A semiconductor quantum dot mimics a two-level atom. Performance as a single photon source is limited by decoherence and dephasing of the optical transition. Even with high quality material at low temperature, the optical linewidths are a factor of two larger than the transform-limit. A major contributor to the inhomogeneous linewdith is the nuclear spin noise. We show here that the nuclear spin noise depends on optical excitation, increasing (decreasing) with increasing resonant laser power for the neutral (charged) exciton. Based on this observation, we discover regimes where we demonstrate transform-limited linewidths on both neutral and charged excitons even when the measurement is performed very slowly.
\end{abstract}

\maketitle

A single quantum dot is a robust, fast, bright and narrow-linewidth emitter of single photons, features not shared by any other emitter \cite{Michler2000,Santori2002,Shields2007}. Future developments in quantum communication place stringent demands on the quality of the photons. For instance, a quantum repeater requires a stream of indistinguishable photons. This can be achieved in a semiconductor only by understanding noise and circumventing its deleterious consequences.

A single quantum dot mimics a two-level atom and single photons are generated either by spontaneous emission from the upper level \cite{Michler2000,Santori2002,Shields2007} or by coherent scattering of a resonant laser \cite{Nguyen2011,Matthiesen2012,Matthiesen2013}. The radiative lifetime is typically $\tau_{R}=800$ ps \cite{Dalgarno2008}. There is evidence that on this timescale, there is negligible pure upper level decoherence provided the quantum dot is at low temperature \cite{Langbein2004,Nguyen2011,Matthiesen2012,Matthiesen2013}. At low Rabi couplings $\Omega$ but higher temperatures (above $\sim 20$ K) \cite{Bayer2002,Kroner2008d}, equivalently at low temperature but at high Rabi couplings \cite{Ramsay2010,Ulrich2011}, phonons dephase the upper level. The remaining issue concerns the wandering of the center frequency over long times \cite{Hogele2004,Houel2012,Prechtel2013}. One way to probe this is with the optical linewidth. Measured on second time-scales, the linewidth $\Gamma$ is typically about a factor of two larger than the transform-limit $\Gamma_{0}=\hbar/\tau_r$ \cite{Hogele2004,Atature2006,Houel2012}, an effect which reduces the indistinguishability of single photons generated far apart in the time domain. 

Both charge noise and spin noise can result in inhomogeneous broadening. Charge noise arises from fluctuations in the electrical environment of the quantum dot, spin noise arises from fluctuations in the nuclear spin ensemble. A diagnostic tool is to add a single electron to the quantum dot. The optical response to charge noise is largely unchanged \cite{Warburton2002} but the response to spin noise is different \cite{Kuhlmann2013b}. The un-paired electron spin in the X$^{1-}$ ground-state splits via the Zeeman effect in the nuclear magnetic field (Overhauser field), Fig.\ 1. Conversely, the X$^{0}$ state is already split at zero magnetic field $B=0$ by electron-hole exchange (the ``fine structure" \cite{Bayer2002}, Fig.\ 1) such that the X$^{0}$ is ``shielded" from the nuclear noise by the hole. In most laser spectroscopy experiments, the linewidths for X$^{0}$ and X$^{1-}$ are very similar \cite{Hogele2004,Houel2012} suggesting that charge noise is responsible for the optical linewidth. This conclusion was questioned recently where strong evidence was presented that in this cold, clean limit, spin noise is responsible for the inhomogenous broadening for both X$^{0}$ and X$^{1-}$ \cite{Kuhlmann2013b,Prechtel2013}.

We show here that spin noise depends sensitively on resonant driving of the optical transition. There is a remarkable dependence on charge. For X$^{0}$, spin noise increases markedly with resonant laser excitation. This increase depends on gate voltage: close to one edge of the Coulomb-blockade plateau \cite{Smith2005}, this ``shake-up"-mechanism can be suppressed and we achieve transform-limited linewidths. Conversely, for X$^{1-}$, resonant optical driving suppresses spin noise. This suppression is effective even without an applied magnetic field and is gate voltage independent. This allows us to demonstrate transform-limited X$^{1-}$ linewidths at modest optical couplings. In both cases, these transform-limited optical linewidths are achieved even when measured on second time-scales. Generally speaking, controlling spin noise is key to operating a quantum dot-based spin qubit  \cite{Merkulov2002,Khaetskii2002,Xu2008,Press2010,Warburton2013}. We show here that controlling nuclear spin noise is also key to creating a quantum dot-based high fidelity single photon source.

\begin{figure}[t]  
\center
\includegraphics[width=\linewidth]{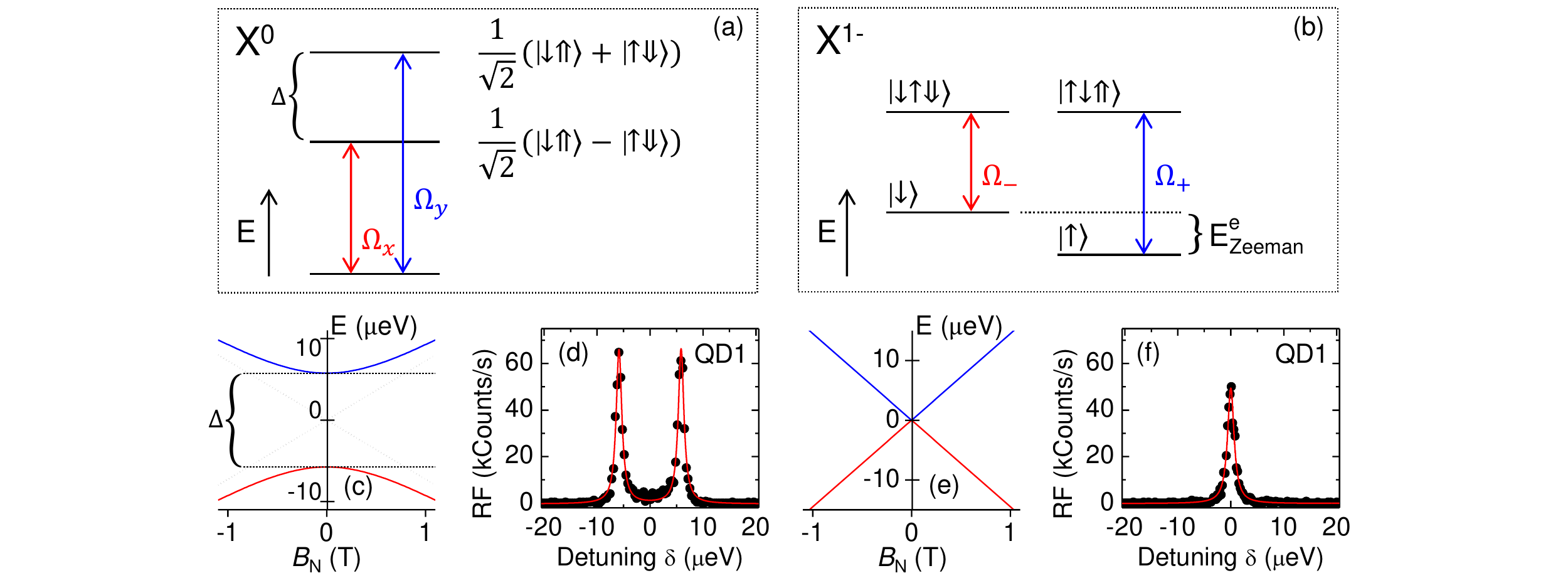}
\caption{(a) Energy levels of neutral exciton X$^{0}$ at zero magnetic field, $B=0$, showing fine structure splitting $\Delta$. (b) Energy levels of charged exciton X$^{1-}$ in an Overhauser field $B_{N}$. (c), (e) X$^{0}$, X$^{1-}$ energy levels versus $B_{N}$ with $\Delta=11.5$ $\mu$eV and electron g-factor $g=-0.5$. (d), (f) X$^{0}$, X$^{1-}$ resonance fluorescence spectra at 4.2 K, $B=0.0$ mT with 100 ms integration time per point. The solid lines are Lorentzian fits to the data. The linewidths are $\Gamma^{X^{0}}=1.29$ $\mu$eV, $\Gamma^{X^{1-}}=1.49$ $\mu$eV; the Rabi energies $\Omega/\Gamma_{0}=0.5$ (X$^{0}$), $0.4$ (X$^{1-}$); and transform-limits $\Gamma_{0}^{X^{0}}=0.92\pm 0.10$ $\mu$eV, $\Gamma_{0}^{X^{1-}}=0.75\pm 0.10$ $\mu$eV.}
\end{figure}

The quantum dots are self-assembled using InGaAs in high purity GaAs and are embedded between a back contact and a surface gate \cite{Supplemental,Houel2012,Kuhlmann2013b}. The gate voltage determines the electron occupation via Coulomb blockade \cite{Warburton2000}. We drive the optical resonance of a single quantum dot at low temperature, 4.2 K, detecting the resonance fluorescence (RF) \cite{Supplemental,KuhlmannRSI2013,Kuhlmann2013b}. The linewidth is determined by sweeping the laser frequency through the resonance, integrating the counts, typically 100 ms per point. $\Gamma_0$ is measured by scanning the optical resonance very quickly such that the fluctuations are frozen \cite{Kuhlmann2013b}, Fig.\ 3(b). A quantum dot noise power spectrum $N_{\rm QD}(f)$ is derived from a Fourier transform of the RF time-trace \cite{Supplemental,Kuhlmann2013b}. From the known relationships between RF signal, Rabi coupling $\Omega$, electric field $F$ and the Overhauser field $B_{N}$, we deduce the variances $F_{\rm rms}$ and $B_{N,{\rm rms}}$ from the noise spectrum \cite{Supplemental,Kuhlmann2013b}.

RF spectra on the neutral, X$^{0}$, and charged, X$^{1-}$, exciton transitions are shown in Fig.\ 1 at $\Omega/\Gamma_{0}=0.5$ (X$^{0}$), $0.4$ (X$^{1-}$). The linewidths are very similar, and are a factor of 1.4 (X$^{0}$), 2.0 (X$^{1-}$) larger than the transform-limit ($\Gamma_0^{X^{0}}=0.92\pm 0.10$ $\mu$eV, $\Gamma_0^{X^{1-}}=0.75\pm 0.10$ $\mu$eV). The increase above the transform-limit represents a sum over all noise sources from the scanning frequency, about 1 Hz, to $\Gamma_{0}$, about 1 GHz. 

\begin{figure}[t]  
\center
\includegraphics[width=\linewidth]{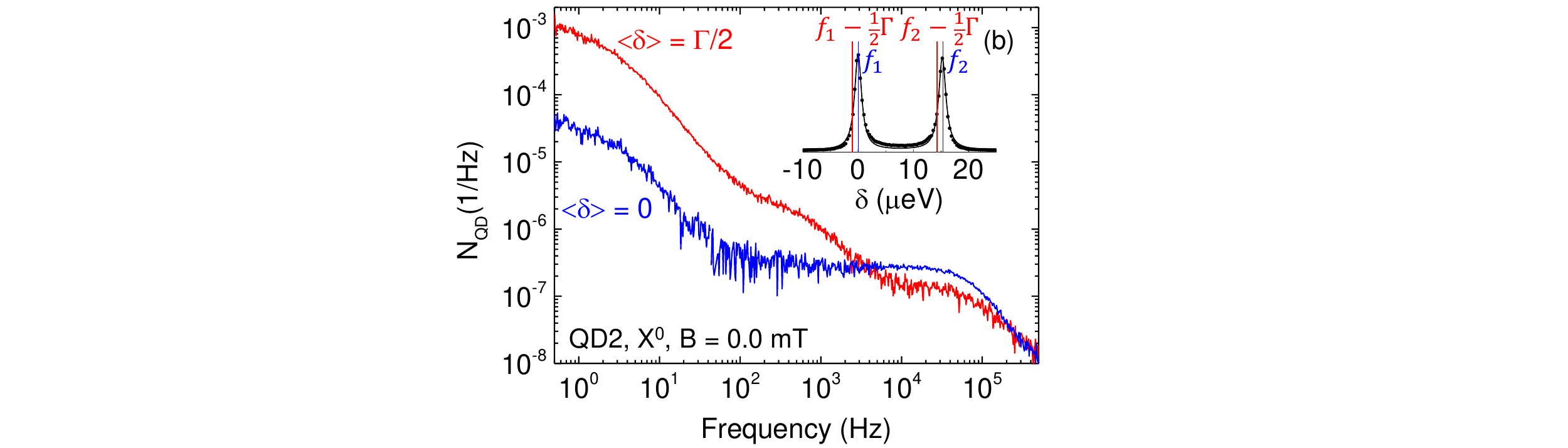}
\caption{ $N_{\rm QD}(f)$ on X$^{0}$ recorded with two lasers of frequencies $f_1$ and $f_2$ and a frequency splitting $f_1-f_2$ equal to the fine structure splitting for $\langle\delta\rangle=0$ (blue) and $\langle\delta\rangle=\Gamma/2$ (red). Inset shows the laser frequency detuning relative to the optical resonance.}
\end{figure}

Noise spectra for X$^{1-}$ and X$^{0}$ are shown in Fig.\ 3(a). In both cases, there is a roll-off feature at low frequencies (linewidth $\sim 30$ Hz) arising from charge noise and a second roll-off feature at higher frequencies (linewidth $\sim 200$ kHz (X$^{0}$), $10$ kHz (X$^{1-}$)) arising from spin noise. The two noise processes can be identified via a dependence on detuning. For X$^{1-}$, the key evidence is the change in $N_{\rm QD}$ at $B=0$ on switching from $\langle\delta\rangle=0$ to $\langle\delta\rangle=\Gamma/2$ which increases/decreases the sensitivity to charge/spin noise \cite{Supplemental,Kuhlmann2013b}. For X$^0$, charge noise moves both peaks in the same direction; spin noise moves them apart or closer together, a ``breathing" motion. A two-laser experiment enables us to distinguish between these two possibilities. Specifically, we record X$^{0}$ noise spectra with two lasers with frequencies separated in frequency by the fine structure. On detuning both lasers from $\delta=0$ to $\delta=\Gamma/2$, the sensitivity to charge noise increases (changing from second order to first order) yet the sensitivity to spin noise decreases, both in exactly the same way as for X$^{1-}$ with one laser \cite{Supplemental}. In the experiment, switching from $\langle\delta\rangle=0$ to $\langle\delta\rangle=\Gamma/2$ causes the noise power of the low frequency component to increase markedly identifying it as charge noise whereas the noise power of the high frequency component decreases, identifying it as spin noise, Fig.\ 2. The two components can be integrated separately: the $f$-sum over the charge noise gives a contribution to $\Gamma$ of $<0.05$ $\mu$eV for both X$^{0}$ and X$^{1-}$ \cite{Supplemental}, a negligible value. We note that both the Stark coefficient and $\Gamma$ vary from quantum dot to quantum dot yet there is no correlation between the two \cite{Supplemental}, pointing also to the unimportance of charge noise in the optical linewidth.

The spin noise spectra yield $B_{N,{\rm rms}}^{X^{0}}=210\pm 20$ mT yet $B_{N,{\rm rms}}^{X^{1-}}=9\pm 3$ mT. Concomitant with the different $B_{N,{\rm rms}}$ values are the associated $B_{N}$-correlation times, much shorter for X$^{0}$ (5 $\mu$s) than for X$^{1-}$ (100 $\mu$s) \cite{Kuhlmann2013b, Stanley2014}. Without optical excitation, $B_{N,{\rm rms}} \sim 20$ mT \cite{Braun2005}, and arises from incomplete cancellation of the hyperfine interaction in the mesoscopic-like nuclear spin ensemble of $N \sim 10^{5}$ nuclei \cite{Merkulov2002,Khaetskii2002}. The transform-limit $\Gamma_0$ is demonstrated for X$^{0}$ and X$^{1-}$ by scanning the resonance at frequencies above $50$ kHz, Fig.\ 3(a,b) \cite{Kuhlmann2013b}. In contrast to the noise spectra, a linewidth measurement at high scanning frequency probes the weakly-excited nuclear spin noise as X$^{0}$ is excited only for a short time. For both X$^{0}$ and X$^{1-}$ the dependence of the optical linewidth on the scanning frequency is Lorentzian with linewidth $30\pm3$ kHz, Fig.\ 3(b), completely consistent with a 100 $\mu$s noise correlation time. This demonstrates that the reduced correlation time and increased amplitude of the spin noise in the X$^0$ noise spectra is related to the constant optical driving.

\begin{figure}[t]  
\center
\includegraphics[width=\linewidth]{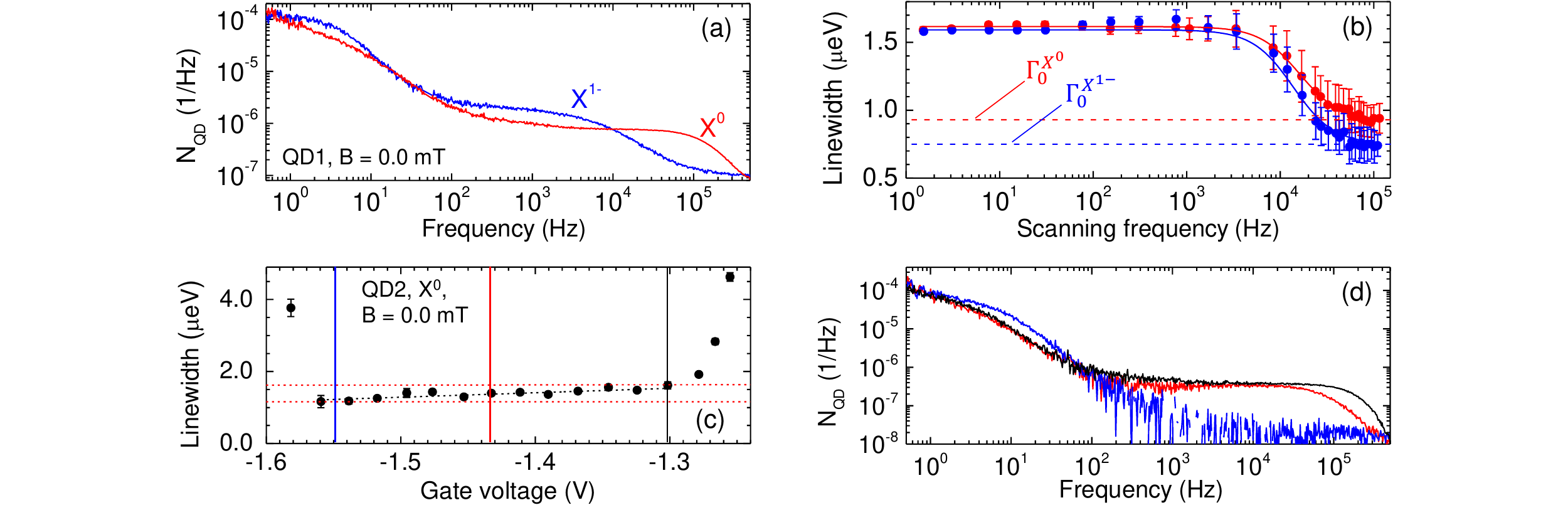}
\caption{(a) $N_{\rm QD}(f)$ for both X$^{0}$ (red) and X$^{1-}$ (blue). (b) RF linewidth against scanning frequency $d\delta/dt/\Gamma_0$. $\Gamma$ approaches $\Gamma_0$ for scanning frequencies above 50 kHz. For each scanning frequency, the error bar represents the standard deviation of several hundred linewidth scans. Solid lines represent a Lorentzian fit of the data with linewidth $30\pm3$ kHz. (c) Optical linewidth measured for different gate voltages by sweeping the laser frequency through the resonance and integrating 100 ms per point. $\Gamma$ decreases from $1.65$ $\mu$eV to $1.16$ $\mu$eV with decreasing gate voltage. (d) X$^{0}$ noise spectra recorded at Rabi energies $\Omega/\Gamma_0=0.65$ for different voltages, indicated in (c) by solid lines. Maximum/minimum spin noise (black/blue) is correlated with the largest/smallest $\Gamma$.}
\end{figure}

The $\Omega$-dependence of $N_{\rm QD}(f)$ is highly revealing, Fig.\ 4. As $\Omega$ increases, the X$^{0}$ spin noise {\em increases}, Fig.\ 4(a). $B_{N,{\rm rms}}^{X^{0}}$ increases roughly linearly with $\Omega$ reaching at the highest couplings extremely high values, 300 mT, Fig.\ 4(b). In complete contrast, the X$^{1-}$ spin noise {\em decreases} as $\Omega$ increases, Fig.\ 4(c), equivalently $B_{N,{\rm rms}}^{X^{1-}}$. ($B_{N,{\rm rms}}^{X^{0}}$ is determined by a Monte Carlo simulation of $N_{\rm QD}(f)$ including an ensemble of fluctuating nuclei -- this is robust as X$^{0}$ is sensitive only to the vertical component of $B_{N}$ \cite{Supplemental}. X$^{1-}$ responds to all three components of $B_{N}$, a more complex problem, and instead $B_{N,{\rm rms}}^{X^{1-}}$ is determined with lower systematic error from the 2-laser experiment.) 

\begin{figure}[t]  
\center
\includegraphics[width=\linewidth]{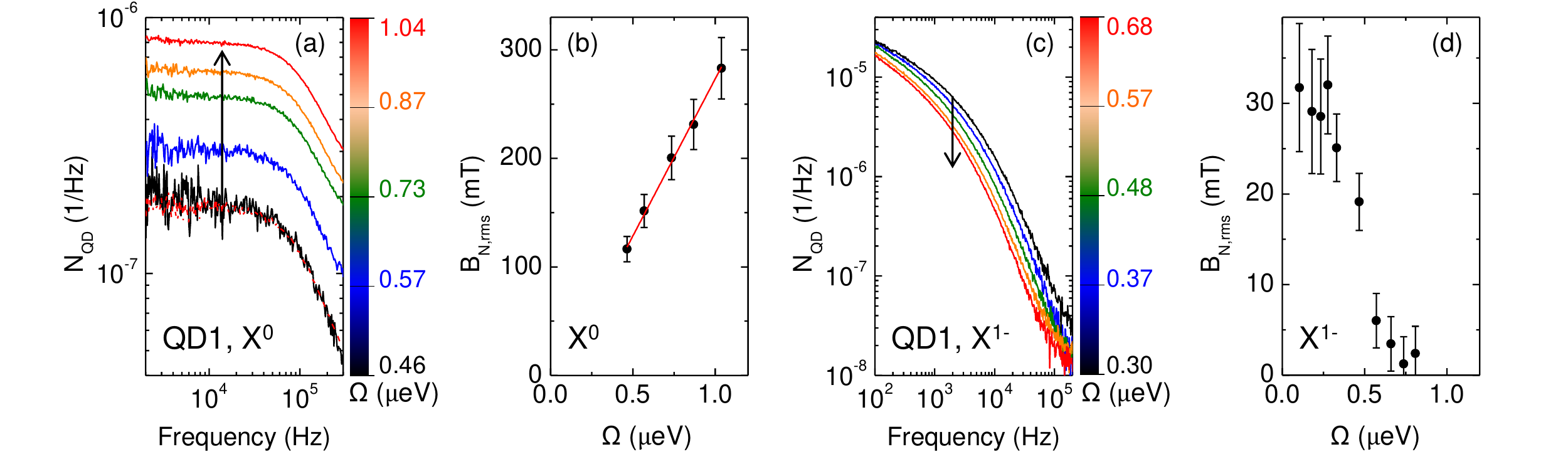}
\caption{(a) $N_{\rm QD}(f)$ on X$^{0}$ for a series of Rabi couplings $\Omega$ at $B=10.0$ mT. The experimental data for $\Omega=0.46$ $\mu$eV (black curve) is accompanied by the Monte Carlo fit (red dashed curve). (c) $N_{\rm QD}(f)$ on X$^{1-}$ for a series of Rabi couplings $\Omega$ (taken at $B=10.0$ mT to enhance the sensitivity to spin noise \cite{Kuhlmann2013b}). (b),(d) $B_{N,{\rm rms}}$ versus $\Omega$ for X$^{0}$, X$^{1-}$.}
\end{figure}

We address whether the spin noise reduction in the case of X$^{1-}$ is sufficient to achieve transform-limited optical linewidths. The $\Omega$-dependence of $\Gamma^{X^{1-}}$ can be described extremely well with the two-level result including an inhomogeneous broadening $\gamma$, Fig.\ 5(b) \cite{Supplemental}. At low $\Omega$, $\Gamma$ is determined by $\Gamma_{0}$ and $\gamma$; at higher $\Omega$, $\Gamma$ increases (``power broadening") and $\gamma$ becomes irrelevant. The solid line in Fig.\ 5(b) therefore represents the ideal limit ($\Gamma$ versus $\Omega$ with $\gamma=0$). A linewidth measurement is complex in the sense that the spin noise is a function of both Rabi energy and detuning. To simplify matters, we performed the experiment with two lasers. The concept is that the stronger, constant frequency pump laser ($\Omega_{2}, \delta_{2}$) determines the spin noise, and the weaker probe laser ($\Omega_{1}, \delta_{1}$) measures the optical linewidth. Fig.\ 5(a) shows $\Gamma^{X^{1-}}$ measured by sweeping $\delta_{1}$ versus $\delta_{2}$ for $\Omega_{1}=0.23,\Omega_{2}=0.80$ $\mu$eV. For large $\delta_{2}$, the pump laser has no effect on $\Gamma$; power broadening is irrelevant and $\Gamma$ is far from the transform-limit. For small $\delta_{2}$ however, $\Gamma$ decreases, despite the power broadening induced by $\Omega_{2}$. Taking into account power broadening, $\Gamma$ reduces to the ideal limit. Fig.\ 5(b) shows the results as $\Omega_{2}$ increases: for $\Omega/\Gamma_{0}>0.75$, transform-limited optical linewidths are achieved.

\begin{figure}[t]  
\center
\includegraphics[width=\linewidth]{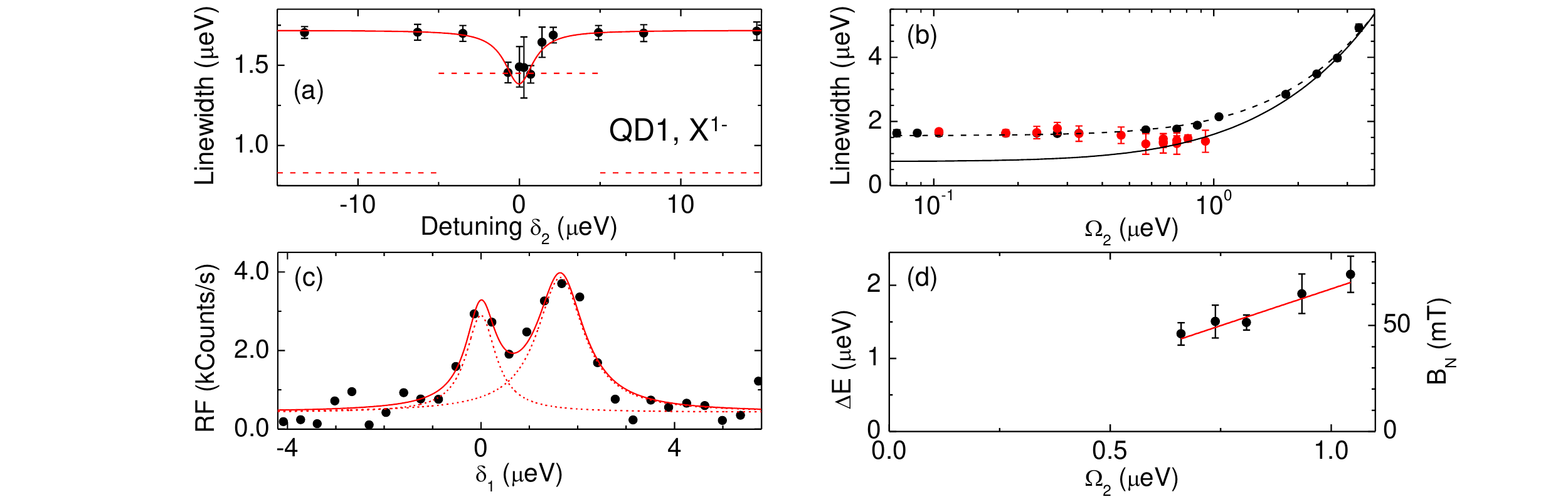}
\caption{The two-laser experiment ($B=0.0$ mT, $T=4.2$ K) on X$^{1-}$. (a) Optical linewidth measured with the probe laser ($\Omega_{1}=0.23$ $\mu$eV) versus detuning of the pump laser $\delta_{2}$ for $\Omega_{2}=0.80$ $\mu$eV. The dashed lines show the ideal case (zero inhomogeneous broadening) in two limits, $\Omega=\Omega_{1}$ and $\Omega=\Omega_{1}+\Omega_{2}$, appropriate for large $\delta_{2}$ and $\delta_{2}=0.0$ $\mu$eV, respectively, the difference arising from power broadening. (b) Optical linewidth in one-laser experiment (black points) versus $\Omega$ with fit to 2-level model ($\gamma=1.35$ $\mu$eV, black curve). The optical linewidth in two-laser experiment ($\Omega_{1}=0.23$ $\mu$eV, $\delta_{2}=0$) versus $\Omega_{2}$ (red points). (c) Probe spectrum with $\Omega_{1}=0.23$ $\mu$eV, $\Omega_{2}=0.80$ $\mu$eV and $\delta_{2}=0.0$ $\mu$eV (points) with a two Lorentzian fit (solid line, energy separation 1.6 $\mu$eV, linewidths $0.8\pm0.3$, $1.2\pm 0.3$ $\mu$eV). (e) Splitting from (d) versus $\Omega_{2}$.}
\end{figure}

The spin noise reduction on driving X$^{1-}$ with the pump laser is accompanied by a profound change in the probe spectrum: the optical resonance now splits into two resonances, Fig.\ 5(c). The splitting reflects a static electron Zeeman splitting in the single electron ground-state, $B_{N}=58$ mT in Fig.\ 5(c), with $B_{N}$ increasing with $\Omega_{2}$, Fig.\ 5(d). Equivalently, even without an applied magnetic field \cite{Chekhovich2010}, a nuclear spin polarization is created by the optical coupling. This demonstrates that the laser locks the nuclear spins into an eigenstate of the $\Sigma I_{z}$ operator.

The large $B_{N,{\rm rms}}^{X^{0}}$ would appear to prohibit transform-limited linewidths on X$^{0}$ at all but the very lowest optical couplings. However, we have discovered that $B_{N,{\rm rms}}^{X^{0}}$ depends not only on optical coupling but also on gate voltage: close to the low-bias edge of the X$^{0}$ Coulomb blockade plateau, both the linewidth (measured slowly with one-laser) and the spin noise (measured with constant driving at detuning zero) decrease, Fig.\ 3(c,d). Additionally. the correlation time increases, Fig.\ 3(d). Closer to the plateau edge, $\Gamma$ rises rapidly on account of strong dephasing via co-tunneling \cite{Smith2005}. However, at the ``sweet-spot",  $B_{N,{\rm rms}}^{X^{0}}$ reduces to $<40$ mT, the linewidth to $\Gamma=1.16 \pm 0.17$ $\mu$eV, which is within error the same as the transform-limit, $\Gamma_0=1.08 \pm 0.10$ $\mu$eV at this coupling (i.e.\ taking into account the small power broadening).

The mechanisms by which spin noise responds to resonant optical excitation are unknown. For X$^{1-}$, the data are compatible with a ``narrowing" of the nuclear spin distribution, perhaps caused by continuous weak measurement via the narrowband laser \cite{Klauser2008}. The correlation time is compatible with the nuclear spin dipole-dipole interaction, but this equivalence is insufficient to make a definite statement. For X$^{0}$ it is unlikely that the standard electron spin-nuclear spin contact hyperfine interaction can offer an explanation, and it is highly unlikely that the bare dipole-dipole interaction can account for the short correlation time. One possibility is that the hole in the X$^{0}$ is important: a hole has a complex hyperfine interaction, containing a term $(I_+ J_z + I_- J_z)$, exactly the structure required to shake-up the nuclear spins on creation of a hole ($I$ is the nuclear spin, $J$ the hole spin) \cite{Ribeiro2014}. While the coefficient of this term is likely to be small, it can have significant consequences should the dark X$^{0}$ state be occupied for times far exceeding the radiative lifetime \cite{Ribeiro2014}. Experimentally, dark X$^{0}$ state occupation is conceivable here, and the dark state lifetime is suppressed at the edges of the Coulomb blockade plateau \cite{Smith2005}, possibly accounting for the observed quenching of the nuclear spin shake-up. We hope that our results will stimulate a refinement in understanding of the exciton-nuclear spin interaction.

In conclusion, we demonstrate the effect of resonant optical excitation on the nuclear spin noise in a single self-assembled quantum dot. Resonant optical excitation decreases nuclear spin noise for a quantum dot occupied with a single electron yet increases the nuclear spin noise for an empty quantum dot. For the empty dot, the nuclear spin shake-up can be suppressed in a specific bias region. Based on these observations, we demonstrate the generation of transform-limited optical linewidths, even when the linewidths are measured very slowly, for both the neutral and charged excitons. It is therefore possible to generate truly indistinguishable photons from the same solid-state emitter even when the photons are created at widely different moments in time.

We acknowledge financial support from NCCR QSIT. We thank Christoph Kloeffel, Daniel Loss, Franziska Maier and Hugo Ribiero for helpful discussions; Sascha Martin and Michael Steinacher for technical support. A.L., D.R. and A.D.W. acknowledge gratefully support from DFG SPP1285 and BMBF QuaHLRep 01BQ1035.\\

\end{document}